\documentclass[12pt]{article}
\usepackage{amsbsy,amssymb}
\newcommand{\fr}{\frac}
\newcommand{\lb}{\label}

\newcommand{\be}{\begin{equation}}
\newcommand{\ee}{\end{equation}}
\newcommand{\ab}{\begin{align}}
\newcommand{\as}{\end{align}}
\newcommand{\abs}{\begin{aligned}}
\newcommand{\ass}{\end{aligned}}
\newcommand{\ba}{\begin{array}}
\newcommand{\ea}{\end{array}}
\newcommand{\beqa}{\begin{eqnarray}}

\newcommand{\la}{\lambda}
\newcommand{\La}{\Lambda}
\newcommand{\del}{\partial}
\newcommand{\eeqa}{\end{eqnarray}}
\newcommand{\ep}{\epsilon}

\newcommand{\Rcc}{\mathbb{R}}

\newcommand{\Omo}{\Omega}

\newcommand{\kd}{\delta}

\begin{document}
\title{}
\author{}
\date{}

\begin{flushright}

hep-th/0001218

\end{flushright}

\begin{center}
{\Large \bf BFV--BRST analysis of  equivalence between 
noncommutative and ordinary gauge theories}

\vspace{1cm}
{\footnotesize \"{O}mer F. DAYI}

\vspace{10pt}

{\footnotesize \it 
Physics Department, Faculty of Science and
Letters, Istanbul Technical University,
80626 Maslak, Istanbul, Turkey 

\vspace{10pt}
and
\vspace{10pt}

Feza G\"{u}rsey Institute,
P.O.Box 6, 81220
\c{C}engelk\"{o}y, Istanbul, Turkey. }\footnote{E-mails:
dayi@itu.edu.tr  and dayi@gursey.gov.tr.} 
\end{center}
\vspace{3cm}

{\small
Constrained hamiltonian structure of noncommutative gauge theory for
the gauge group $U(1)$ is discussed. Constraints are shown to
be first class,
although, they do not give an Abelian algebra in terms of  Poisson
brackets. The related BFV-BRST charge
gives a vanishing generalized Poisson bracket by itself   due to
the associativity of
 $*$--product. Equivalence of noncommutative and ordinary
gauge
theories is formulated in generalized phase space by using 
BFV-BRST
charge and a solution is obtained. Gauge fixing is  discussed.
}

\vspace{5cm}

\pagebreak

Noncommutative geometry is one of the most attractive subjects of physics
and mathematics\cite{acb}. 
Attention to it is increased considerably
after the observation that in
string theories noncommutativity of space appears in a natural
way\cite{nis}.
Seiberg and Witten used the ideas about the noncommutativity
in string theory and showed that 
 noncommutative and ordinary gauge theories are equivalent\cite{sw}.
Following them there appeared 
several works  on noncommutative gauge theories\cite{arw}.   
BRST symmetry of  a
noncommutative gauge theory in first order Lagrangian
 formulation was discussed in \cite{ben}.

When one deals with a gauge theory, Lagrangian framework is suitable for
perturbative calculations. However, 
understanding its
Hamiltonian structure is 
essential  to perform  canonical quantization which can be used to
derive some features like the correct measure of  path integral
and  physical states.

In Hamiltonian framework,
an ordinary gauge theory action leads to first class constraints
which decrease the number of linearly independent phase space variables.
For keeping track of  gauge invariance, instead of getting rid of the
unphysical degrees of freedom, one enlarges the phase space by introducing
ghost fields possessing the opposite statistics of the constraints and
write the BVF--BRST charge\cite{bfv}.
Now,  invariance of the theory under the transformation of variables
generated by
the constraints (reminiscent of gauge invariance) replaced by the
invariance
under the transformations generated by the BFV--BRST charge.
Action and measure of the related path integrals 
should be invariant under the BFV--BRST transformations.
Once the canonical commutation relations between  phase space fields
are imposed,
they can be written in terms of their normal modes. Thus, one can find
the quantum BFV--BRST charge which is defined to be nilpotent and whose
cohomology gives the physical states\cite{ht}.

One expects that canonical formulation of noncommutative gauge theories
can be studied in a similar manner, as far as the time coordinate is
kept classical. We show that, indeed, this is the case, although,
it was not guaranteed a priori: Gauge invariance of  noncommutative gauge
theory is not an invariance of the Lagrange density, 
in contrary to the ordinary case, but  an invariance of the action,

Hamiltonian structure of
ordinary gauge theories are well studied.
Thus, we hope that  showing the
equivalence between ordinary and noncommutative gauge theories
in terms of  the generalized phase space variables
will be useful in understanding features like  canonical quantization
of the latter.

Constrained hamiltonian  structure of noncommutative gauge
theory for the gauge group $U(1)$ is studied.
This is the simplest gauge group, however, the main features of
noncommuting gauge theories are already present. 
The related BFV-BRST charge which
gives a vanishing generalized Poisson bracket with itself
is presented.
Equivalence of ordinary and
noncommutative gauge theories is formulated in terms of
generalized phase
space variables and a solution is given. 
We also briefly discuss gauge
fixing within this formalism.

Noncommutative gauge theory for $U(1)$ in $\Rcc^d$
is given with the action
\be
\lb{nyma}
S=-\fr{1}{4} \int d^dx\ F_{\mu\nu}* F^{\mu\nu}
\ee
in terms of the metric $\eta_{\mu\nu} ={\rm diag} (-1,1,\cdots ),$
and 
\[
F^{\mu\nu}(x)=\del^\mu A^\nu -\del^\nu A^\mu
-i\left( A^\mu * A^\nu -A^\nu * A^\mu \right) .
\]

$*$--product is defined as
\be
G(x)* K(x) = G(x) e^{\fr{i}{2}\theta^{ij}
\stackrel{\leftarrow}{\partial_i}
\stackrel{\rightarrow}{\partial_j} }
 K(x) ,
\ee
where
$\theta^{ij}$ is an antisymmetric, constant matrix.
The $*$--product reflects the fact that
the space coordinates $x_i$
are noncommuting:
\[
x^i*x^j-x^j*x^i =i\theta^{ij} .
\]

(\ref{nyma}) is invariant under the gauge transformations
\[
\delta A_\mu  = \del_\mu \la  +i(\la * A_\mu 
-A_\mu * \la ),
\]
because we have the equalities
\beqa
\int d^{d-1}x\  G(x)* \left( K(x)* L(x)\right)
& = & \int d^{d-1}x\   G(x) \left(K(x)* L(x)\right) \nonumber \\
& = & \int d^{d-1}x\   \left( L(x)*G(x)\right) K(x), \nonumber \\
& = & \int d^{d-1}x\   \left( G(x)*K(x)\right) L(x), \nonumber
\eeqa
which follow from the assumption that all of the fields which we deal
with are vanishing at infinity.

The time coordinate is not deformed and 
noncommutativity is due to the $*$--product. Hence, the fields are
ordinary ones and the canonical momenta are defined as usual:
\beqa
P_0=\fr{\del S}{\del\dot{A}^0} & = & 0,  \lb{pco} \\
P_i=\fr{\del S}{\del \dot{A}^i} & = & \del_0 A_i 
-\del_i A_0 -i(A_0* A_i -A_i * A_0).
\eeqa
The canonical hamiltonian reads
\be
\lb{ch}
H=\int d^{d-1}x\ \left( \fr{1}{2} P^2_i +\fr{1}{4} F_{ij} * F^{ij} 
-A_0 \Phi (x) \right) ,
\ee
where 
\be
\lb{con}
\Phi (x) =\del_i P^i +i[P_i,A^i],
\ee
written in terms of the Moyal bracket 
\[
[G(x),K(x)]\equiv G(x)* K(x) -K(x)* G(x).
\]
The primary constraints (\ref{pco}) should be constant in time:
\be
\lb{vp}
\{P_0 (x) ,H \}=0,
\ee
where the basic 
Poisson brackets are
\[
\{P_\mu (x), A_\nu (y) \} =\eta_{\mu\nu}\kd (x-y).
\]
The condition (\ref{vp})
leads to the secondary constraints
\be
\lb{Phi}
\Phi(x) =0.
\ee

Obviously, $P_0$ gives a vanishing Poisson bracket with $\Phi(x).$
To classify the constraints $\Phi(x)$ let us introduce a bosonic
parameter $\la (x)$ and deal with 
\beqa
\Phi_\la & \equiv & 
 \int d^{d-1}x\  \Phi (x) \la (x) \nonumber \\
 & = &  \int d^{d-1}x\  \Phi (x) * \la (x) \nonumber \\
 &= &  \int d^{d-1}x\  P_i \left(-\del_i  \la (x)
+i [P_i(x),\la (x)] \right). \nonumber 
\eeqa
The last equality follows from the fact that the Moyal bracket
in (\ref{con})
possesses only odd powers of $\theta^{ij}.$
Poisson bracket of the integrated constraints  is calculated:
\be
\{ \Phi_\la ,\Phi_\kappa \} =i \Phi_{[\la ,\kappa ]},
\ee
where on the right hand side parameter is the Moyal bracket of the ones
appearing on the left.
Moreover, $\Phi (x)$ do not lead to new constraints:
\[
\{H_0, \Phi_\la \} =0,
\]
where $H_0=H|_{\Phi =0}.$
Hence the whole set of constraints is given by
 (\ref{pco}) and (\ref{con}) 
which are first class.

For the following discussions  the primary constraints
(\ref{pco})
 are not essential. Hence, we do not deal with them any more by
fixing the gauge as $A_0=0$ and setting \mbox{$P_0 (x)=0.$} 

To perform the BFV-BRST analysis, we
enlarge the phase space with 
the anticommuting ghost
fileds $C(x)$ and their canonical conjugates $P_C(x)$ 
and use the generalized Poisson bracket
\be
\lb{gcg}
\{ G,K \} =\int d^{d-1}x\ \left( 
\fr{\del_r G}{\del {\cal P} (x)}   \fr{\del_l K}{\del  {\cal Q}(x)} 
-(-1)^{\eta (G) \eta (K)}
\fr{\del_r K}{\del  {\cal P}}
 \fr{\del_l G}{\del {\cal Q} } \right) .
\ee
Here the
subscripts $r$ and $l$ indicate right and left derivatives,
$\eta (G)$ Grassmann parity of $G$ and ${\cal Q},\ {\cal P}$ denote
the generalized phase space coordinates and momenta including the
original ones and the ghosts. We attribute also ghost numbers 
\[
{\rm gh} (C)=-
{\rm gh} (P_C)=1\ ,\ 
{\rm gh} (A_i)=
{\rm gh} (P_i)=0.
\]
The BFV-BRST charge is
\beqa
\Omo & = &
\int d^{d-1}x\  
\left( \Phi (x)* C(x) - P_C(x) \left( C (x) 
* C(x)\right) \right)\nonumber\\
& = & \int d^{d-1}x\  \left( \Phi (x) C(x) - \fr{1}{2}(P_C(x) * C (x)
+ C (x)* P_C )C(x) \lb{Omo}
\right).
\eeqa
To deal with  fermionic fields,
we generalize the Moyal bracket as
\[
[G,K] \equiv G* K - (-)^{\eta (G) \eta (K)} K* G.
\]
Obviously, the generalized Moyal brackets satisfy the generalized
Jacobi identity. Thus,
\[
\int d^{d-1} x\ ( C * C ) [P_C,C]
=\fr{1}{3}\int d^{d-1} x\ \left(
[[C,C],P_C]+[[C,P_C],C]+[[P_C,C],C]
\right) C =0,
\]
which  is the unique nontrivial term to conclude that the BFV-BRST charge
satisfies
\be
\{\Omo ,\Omo \} =0.
\ee

By introducing the rigid, fermionic parameter $\ep$ 
possessing ghost number $-1,$ the phase
space fields transform as
\beqa
\delta_\ep A_i & =& -\ep\left( \del_i C+iC*A_i-iA_i*C\right) , \lb{c11}\\
 \delta_\ep C & = & -i\ep\  C*C\ ,\\
\delta_\ep P_i & = & i \ep \left( P_i * C -C * P_i \right),\\
 \delta_\ep P_C & = & - \ep  \left( \Phi +iP_C * C +iC * P_C  \right)
\lb{c14},
\eeqa
under the BFV-BRST charge (\ref{Omo}).

Now, we would like to
discuss the equivalence of noncommutative and ordinary gauge theories. 
Ordinary gauge theory  action for the gauge group $U(1)$
is given in terms of the field strength
$ f_{\mu \nu }=\del_\mu a_\nu -\del_\nu a_\mu,$ as 
\be
S_o=\int d^d x\  f_{\mu\nu}f^{\mu \nu}.
\ee
The definition of canonical momenta $p_\mu ,$  yields 
the primary constraints $p_0(x)=0.$ These lead to the secondary  
constraints
$\del_ip_i=0,$ which are first class.
We set strongly $p_0=0$ by fixing the gauge as $a_0=0.$
 Now, as usual we enlarge the phase
space with the anticommuting ghost field $c(x)$ and its canonical
conjugate $p_c(x),$
possessing ghost numbers $1$ and $-1$ 
to write the BFV-BRST charge as 
\be
\Omo^o =\int d^{d-1}x\  \del_ip^i c.
\ee
Thus the BFV-BRST transformation of the fields  is given  as
\beqa
\delta^o_\ep a_i =-\ep \del_i c &,& \delta^o_\ep c=0, \lb{c21} \\
\delta^o_\ep p_i = 0 &,& \delta^o_\ep p_c= -\ep \del_i p_i. \lb{c24}
\eeqa

Let us denote the phase space variables in a unified notation as
\be
Q_z \equiv (A_i,P_i,C,P_C)\ ,\ 
q_z\equiv (a_i,p_i,c,p_c).
\ee
Then, we can formulate
equivalence 
of the noncommutative and
ordinary
gauge theories as
\be
\lb{eqr}
Q_z (q)+\delta_\ep Q_z(q) =Q_z(q+\kd^o_\ep q).
\ee
Moreover, there are the conditions
\be
\lb{eqrr}
Q_z(q)|_{\theta =0}=q_z.
\ee
To first order in $\theta^{kl},$
there is a solution of (\ref{eqr})--(\ref{eqrr}): 
\beqa
A_i(a) & = &
a_i -\theta^{kl}(a_k\del_l a_i -\fr{1}{2}a_k\del_ia_l),\lb{s1} \\
P_i(a,p) & = & p_i-\theta^{kl} a_k \del_l p_i, \lb{s2} \\
C(a,c) & = & c+\fr{1}{2}\theta^{kl}\del_k c a_l \lb{s3} \\
P_C(a,p,p_c) & = & p_c +p_c(\del_j p^j)^{-1}\theta^{kl} \del_kp_if_{il}
+\theta^{kl}\del_kp_ca_l. \lb{s4}
\eeqa
As expected, the  solutions  $A_i(a)$  (\ref{s1}) and $C(a,c)$ (\ref{s3})
can be derived by using the solution for $A_i$ and the gauge
parameter $\la$ given in \cite{sw}.
At first glance  (\ref{s4}) can be 
thought of being
inconsistent with the
constraint
$\del_ip^i=0$ of the ordinary $U(1)$ gauge theory.
However, this is due to the fact that noncommutative and
ordinary gauge groups can not be isomorphic to each other\cite{sw}:
when one deals with the noncommutative gauge theory the constraint
$\Phi=0$ leads to $\theta^{ij}=0$ for $\del_ip^i=0$ and  arbitrary
gauge field $a_i.$ There is another way of explaining this fact: observe
that the BRST transformations of $P_C$ and $p_c$ (\ref{c14}),
 (\ref{c24}) are
proportional to the related  constraints, so that they are aware
of the structure of the gauge group. Thus,
as far as we
deal with nonvanishing $\theta^{ij}$ we assume that
 $\del_ip^i\neq 0.$

For performing  gauge fixing one enlarges the phase space with
the bosonic and fermionic canonical conjugate pairs 
$(\La ,\ \Pi )$ and $(\bar{C} ,\ \bar{P}_C)$ satisfying
the generalized Poisson bracket relations
\beqa
\{ \La (x),\Pi (y)\}=\kd (x-y) & ; &
 \{ \bar{C}(x),\bar{P}_C (y)\} =\kd (x-y).
\eeqa
$\La ,\ \Pi$ possess zero  ghost number and 
\[
{\rm gh} (\bar{C})=-{\rm gh} (\bar{P}_C)=-1.
\]
Gauge fixed action which can be used in the related 
path integral can be given in terms of the gauge fixed hamiltonian
\[
H= H_0 +\{ \Psi , \Omo^\prime \},
\]
where $\Psi$ is the gauge fixing fermion possessing ghost number $-1$
and 
\[
\Omo^\prime =\Omo + \int d^{d-1}x\  \Pi \bar{P}_C.
\]

The easist choice for the gauge fixing fermion $\Psi$ is 
\be
\Psi=\int d^{d-1}\ \left( P_C\xi +\bar{C}\La \right) ,
\ee
where $\xi$ is a function of the original fileds and
 defined to possess  a
nonvanishing Poisson bracket with $\Phi (x) .$ The gauge fixed
hamiltonian in this gauge becomes
\be
H=H_0+\int d^{d-1}x\ \left( C\{ \Phi , \xi \} \bar{C}+
\La \left( \Phi +[C,P_C] \right)+\bar{P}_CP_C+\Pi\xi \right) .
\ee

By choosing a gauge condition $\xi (x)$ we can perform perturbative
calculations within the Hamiltonian approach
 to  study noncommutative gauge theory and
its relations
with ordinary gauge theory by making use the
solutions (\ref{s1})--(\ref{s4}). 

By using the constrained Hamiltonian structure presented, one can also
proceed to perform operator quantization  by
expanding the generalized phase space variables in terms of normal modes.


\newpage

\begin{thebibliography}{99}

\bibitem{acb}A. Connes, Noncommutative Geometry, Academic Press, London, 
1994.

\bibitem{nis}A. Connes, M. R. Douglas and A. Schwarz, JHEP 
9802  (1998) 003 .

\bibitem{sw}N. Seiberg and E. Witten, JHEP 9909 (1999) 032.

\bibitem{arw}L. Cornalba, D--brane physics and noncommutative Yang--Mills
theory,  hep-th/9909081;\\
N. Ishibashi, A relation between commutative and
noncommutative descriptions of D--branes, hep-th/9909176;\\
K. Okuyama, JHEP 0003 (2000) 016;\\
T. Asakawa and I. Kishimoto, JHEP 9911 (1999) 024;\\
B. Jurco and P. Schupp, Noncommutative Yang--Mills from
equivalence of star products, hep-th/0001032;\\
S. Terashima, On the equivalence between noncommutative
and ordinary gauge theories, hep-th/0001111;\\
J. Madore, S. Schraml, P. Schupp and J. Wess, Gauge theory
on noncommutative spaces, hep-th/0001203.

\bibitem{ben}H. B. Benaoum, Perturbative BF Yang--Mills
theory on noncommutative $\Rcc^4$, hep-th/9912036.

\bibitem{bfv}E. S. Fradkin and G. A. Vilkovisky, Phys. Lett. B 55 (1975) 
224;\\
I. A. Batalin and G. A. Vilkovisky, Phys. Lett. B 69 (1977) 309;\\
I. A. Batalin and E. S. Fradkin, Phys. Lett. B 122 (1983) 157. 

\bibitem{ht}M. Henneaux and C. Teitelboim,  Quantization of
Gauge Systems, Princeton U. Press, Princeton, 1992.

\end{thebibliography}
\end{document}